\documentclass[twocolumn]{svjour3}          
\RequirePackage{fix-cm}
\RequirePackage{amsmath,amssymb}
\pdfoutput= 1
\usepackage{graphicx}
\usepackage{natbib}

\usepackage[draft]{hyperref}
\usepackage[usenames]{color}

\journalname{Journal of Computational Neuroscience}
\begin{document}

\title{Feedforward Architectures Driven by Inhibitory Interactions
}


\author{Yazan N. Billeh        \and
        Michael T. Schaub}


\institute{Y. N. Billeh \at
			  Computation and Neural Systems Program,
			  California Institute of Technology
              \textit{Current Address:} Allen Institute for Brain Science,
              Seattle.
              \email{yazanb@alleninstitute.org}           
           \and
           M. T. Schaub \at
           ICTEAM, Universit\'e catholique de Louvain;
           Institute for Data, Systems, and Society,
              Massachusetts Institute of Technology;
              Department of Engineering Science, University of Oxford
              \email{mschaub@mit.edu}
}


\maketitle

\begin{abstract}
     Directed information transmission is \newline paramount for many social, physical, and biological systems. 
     For neural systems, scientists have studied this problem under the paradigm of feedforward networks for decades. 
     In most models of feedforward networks, activity is exclusively driven by excitatory neurons and the wiring patterns between them, while inhibitory neurons play only a stabilizing role for the network dynamics.  
     Motivated by recent experimental discoveries of hippocampal circuitry, cortical circuitry, and the diversity of inhibitory neurons throughout the brain, here we illustrate that one can construct such networks even if the connectivity between the excitatory units in the system remains random. 
     This is achieved by endowing inhibitory nodes with a more active role in the network. 
     Our findings demonstrate that apparent feedforward activity can be caused by a much broader network-architectural basis than often assumed.
\end{abstract}

\section{Introduction}
The ability to reliably propagate signals in a targeted manner is essential for the operation of many natural systems, and a necessary building block to establish many further computational mechanisms.
Prototypical models for such targeted information transmission within a neural substrate are feedforward networks, which have been considered in the literature for decades.
In these models, the basic paradigm is to group nodes (neurons) into separate layers, each of which receives excitatory input from the preceding layer, and projects excitatory connections to the subsequent layer (see the reviews~\citet{Vogels2005a,Kumar2010}).
The thus established forward-directed excitatory pathways guide the activity sequentially through the layers.
A large number of variations of this scheme have been considered, such as embedding excitatory feedforward architectures in randomly connected networks to examine their effect on the overall network dynamics and signal propagation~\citep{Mehring2003,Kumar2008}.
Further, feedforward networks have been shown to propagate firing rates~\citep{Rossum2002,Vogels2005},  synchrony/pulse packets~\citep{Diesmann1999,Cateau2001,Kistler2002,Litvak2003,Aertsen1996,Gewaltig2001}, combinations of firing rates and synchronous spiking~\citep{Kumar2010}, and even the ability to gate activity transmission~\citep{Vogels2009}.  
However, within this paradigm of feedforward networks, inhibitory units (neurons) play merely a balancing role: they ensure that the network remains stable, either separately for every layer or globally.

Interestingly, recent work in neuroanatomy has revealed an enormous diversity of inhibitory neurons \newline\citep{Kepecs2014,Klausberger2008,Isaacson2011,Roux2015,Harris2015,Huang2014,Taniguchi2014,Olsen2012,Bortone2014}.
Moreover, specific plasticity rules for different subtypes of inhibitory neurons~\citep{Chen2015} add further to their diverse and heterogeneous connectivity profiles.
In this light it would be surprising if inhibitory neurons serve the cortex only in a homogeneous, passive role when it comes to information propagation, as is assumed in most feedforward networks.

While some studies exist that specifically account for intra- and interlayer inhibition \citep{Aviel2003,Tetzlaff2003,Mehring2003,Vogels2005,Kremkow2010}, information is still mediated by a cascade of excitatory neurons.
However, as experimental evidence suggests, the nervous system likely uses a combination of methods to transmit information \citep{Reyes2003,Vincent2012}.
An important natural question is thus if embedded excitatory pathways are necessary for feedforward propagation or whether one can construct networks in which there are no preferred excitatory-to-excitatory pathways, but \textit{inhibitory neurons} play a pivotal role for the propagation of activity between layers.
In the following we demonstrate that such feedforward processing is indeed possible with two exemplary circuits.

\begin{figure*}[tb!]
  \centering
  \includegraphics{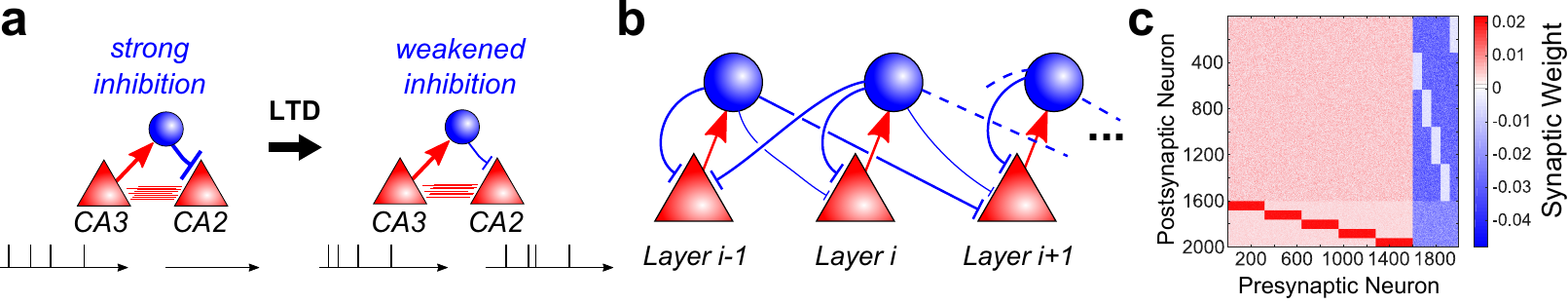}
  \caption{\textbf{Cross coupled feedforward networks.}
      \textbf{(a)}~Schematic of the finding of Ref.~\citep{Nasrallah2015}. Inhibitory long term depression (LTD) lowers the feedforward inhibition of CA2, allowing information transfer from CA3 to CA2.
      \textbf{(b)} Schematic of proposed network architecture, which results from cascading the post-LTD connection motif in (a) in a chain-like manner.
      Note that only connections that are not identically distributed in the rest of the network are displayed for visual clarity. 
      The feedback between excitatory and inhibitory neurons drive the feedforward activity, while connections between alike neurons remain uniform (see text).
      \textbf{(c)} Synaptic weight matrix of an example network with five groups. Note that the network contains connections between all neuron types. Panel \textbf{b} only emphasizes the dominant pathways altered in the ccFFN architecture.
      Finally, note that we do not model plasticity in our networks and use fixed weight matrices only.}
  \label{fig:1}
\end{figure*}

\section{Results}

\subsection{Feedforward networks and information transfer in neural networks}
A core question in systems neuroscience is how computations are performed within a neural substrate.
Such computations may include a variety of things, such as eliciting a downstream response to a sensory input, or some forms of integration or aggregation of information.
From a mathematical viewpoint, we may view such a computation as performing a certain transformation $f$ of an input signal $x$ to an output signal $y$, where $x$ and $y$ may be defined in a context-specific way, e.g., in terms of firing rates, or spike timings of a particular subset of neurons.
This viewpoint immediately triggers a certain set of questions, such as: what type of functions can be implemented? 
Given that there is noise in the system, how reliable can these transformations be performed? 
How fast and how efficient can such operations be performed?

While feedforward propagation may seem like a quite simple function from an input-output perspective, as it simply amounts to a (time-delayed) identity mapping, i.e., the reproduction of the `input information' at another location (or layer), the inherent noise in neural systems makes reliable propagation non-trivial to implement.
More importantly, however, feedforward propagation forms a primitive for gating, multiplexing or other more complex signal transformations.
Hence, feedforward propagation in neural networks has been subject to intense study in the past.

The primary focus of such studies has been on characterizing pathways of excitatory neurons, and the fidelity of the input/output mapping of such excitatory feedforward networks.
In the following we will be less concerned with a precise input-output characterization of the system, but more with its architecture.
Namely, we are primarily interested in how far apparent feedforward propagation, as observed in a rastergram, may be implemented by different neural architectures, which do not rely on the propagation of information via a dominant excitatory pathway.
Conceptually, this implies that the structural repertoire to implement this kind of behavior is larger than one may naively assume; a lack of directional preference in the excitatory-to-exctitatory coupling between neurons, does not imply a lack of directional information propagation. 
A corollary of this finding is that the observation of a `feedforward like' pattern in a rastergram does not implicate that a strong directional excitatory pathway is present between the source and the target neurons.

\subsection{Cross-coupled feedforward networks}
The high diversity of inhibitory neuron subtypes \citep{Kepecs2014,Klausberger2008,Isaacson2011,Roux2015,Harris2015,Huang2014,Taniguchi2014,Olsen2012,Bortone2014}, the different plasticity behaviors \citep{Chen2015}, and the increasing list of discoveries of long-range GABAergic neurons~\citep{Toth1993,Freund1988,Acsady2000,Gulyas1990,Rocamora1996,Freund1991,Freund1992,Pinto2006,Melzer2017} heavily suggests that a range of roles are played by inhibitory neurons in neural computations.
While the network architectures we introduce below are topological, and do not imply any physical distance constraints, our introductory example is nevertheless motivated by a recent experimental study of the hippocampus~\citep{Nasrallah2015} where it was demonstrated that excitatory neurons in the CA3 region are unable to drive excitatory neurons in area CA2 due to strong feedforward inhibition.
However, when this strong inhibition of CA2 is alleviated, CA3 can indeed excite CA2 excitatory cells to elicit action potentials~\citep{Nasrallah2015} (Fig.~\ref{fig:1}a). 
An interesting feature of this finding is that the directionality in the interaction between CA2 and CA3 appears to be dictated by the connections between excitatory and inhibitory neurons, rather than a consequence of unidirectional excitatory connections targeting CA2. 
Indeed, excitatory connections between CA2 and CA3 are reciprocal~\citep{Kohara2014,Llorens-Martin2014} -- yet there is still a \textit{directed} propagation towards CA2~\citep{Nasrallah2015}.
Stated differently, the targeted activation of CA3 is controlled by an excitatory-inhibitory-excitatory pathway.

We sought to leverage this targeted activation mechanism by cascading this connection motif using leaky-integrate-and-fire (LIF) networks (see Materials and Methods).
The result is a circuit with \textit{uniform} connectivity among excitatory neurons with \textit{no preferred direction}, which is nevertheless able to elicit feedforward activity due to the specific cross coupling between the excitatory and inhibitory neurons.
We term this circuitry a cross-coupled feedforward network (ccFFN). 
A schematic can be found in Fig.~\ref{fig:1}b that emphasizes the main properties of the network without including all the connections.
To give a more detailed picture, Fig.~\ref{fig:1}c displays the weight matrix of a network instantiation with 5 groups (with the last group connecting to the first) which illustrates the full connectivity of ccFFNs.

The behavior of ccFFNs can be explained by the following rationale: 
(i) the excitatory neurons in each layer are more strongly coupled to the group of inhibitory neurons in their own layer relative to other inhibitory neurons; 
(ii) an activity increase of such an excitatory group thus triggers elevated activity in the corresponding inhibitory neurons; 
(iii) this inhibitory group of neurons targets the subsequent layer of excitatory neurons \emph{more weakly} relative to other excitatory neurons; 
(iv) the reduced inhibition (relative) of the subsequent excitatory group leads to increased excitatory activity in the subsequent layer, while the activity in the initial layer returns to baseline; 
(v) by cascading this cross-coupling motif, elevated activity of excitatory neuron groups propagates through the circuit. 

\begin{figure*}[tb!]
  \centering
  \includegraphics{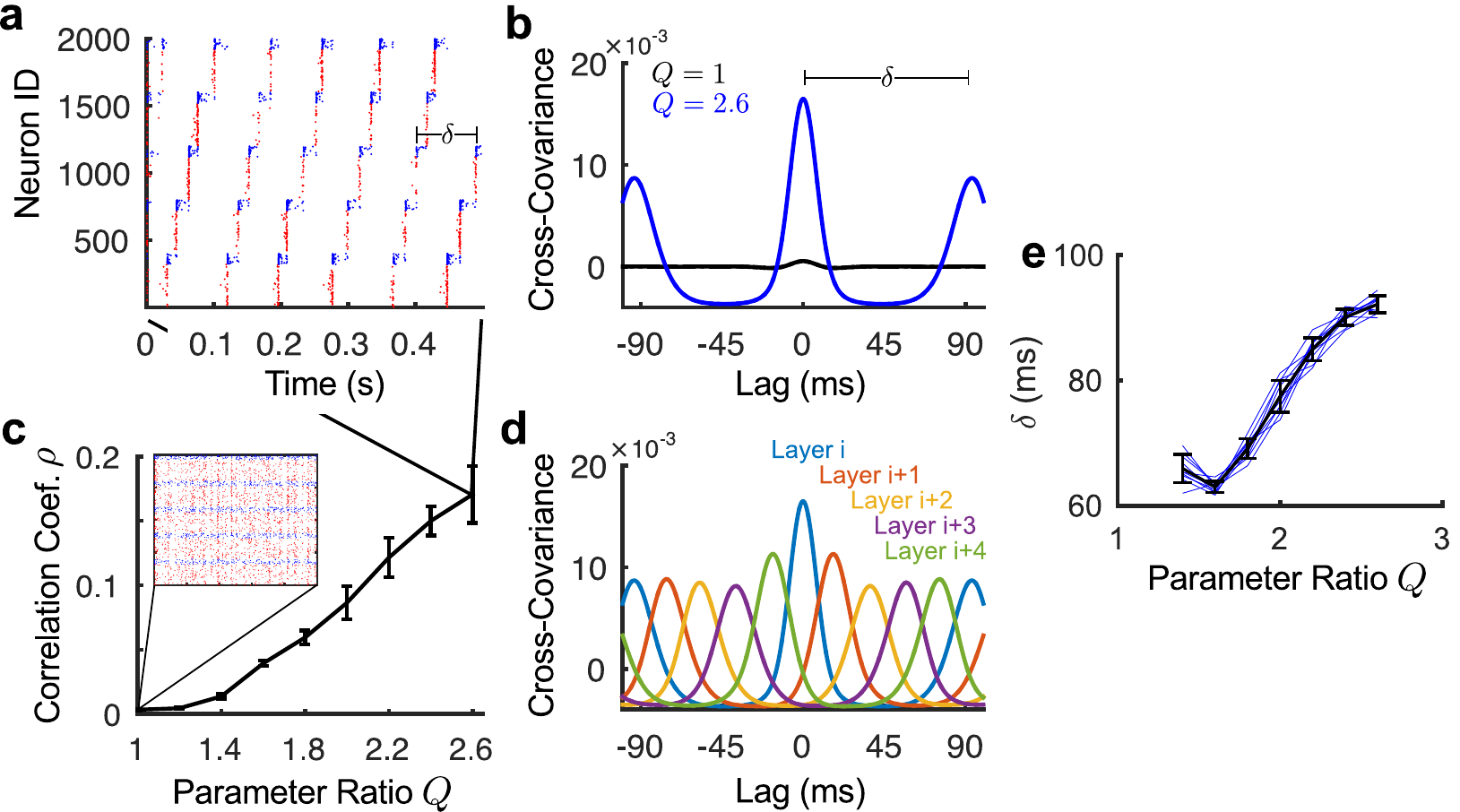}
  \caption{\textbf{Firing pattern characterization of cross coupled feedforward networks}. 
      \textbf{(a)} Example raster plot with 5 groups that show the propagation of activity between layers. 
      Observe that inhibitory neurons also show feedforward propagating activity and that the final group connects back to the first (circular arrangement) and hence the activity propagates indefinitely.
      \textbf{(b)} Group-averaged cross-covariance for parameter ratio values $Q = 1$ and $Q = 2.6$.
      For $Q=1$ (no feedforward structure) the firing is clearly not synchronous and does not display any pattern.
      In contrast, with imposed cross-coupled feedforward structure ($Q=2.6$), there is a peak indicating strong synchronous firing inside each layer.
      The second peak at time $\pm \delta$ indicates the periodically repeating firing pattern (see also \textbf{(c)}).
      \textbf{(c)} The average Pearson correlation coefficient within layers as a function of $Q$.
      The larger the feedforward ratio, $Q$, the higher the correlation of firing within layers.
      Error bars are standard deviations.      
      \textbf{(d)} Average cross-covariance of the neural firing patterns in layer $i$ with neurons in all other layers for $Q=2.6$. 
      Observe that the time lag of the peaks are arranged consecutively, illustrating the orderly feedforward progression between groups.
      \textbf{(e)} Plot of the firing time period $\delta$, as shown in (a), (b), as a function of $Q$.
      Note that for very small $Q$, the cross-covariance would have no secondary peaks (see (b)).
      We thus imposed a threshold that required the secondary peak to be at least half as large as the primary peak.
  We only plot $\delta$ values where this condition was satisfied (starting from $Q=1.4$).}
  \label{fig:2}
\end{figure*}

A simulation of a network with such a ccFFN topology is shown in Fig.~\ref{fig:2}a.
Note that, to eliminate transient effects from particular driving inputs we connected the last layer with the first layer, thus establishing a circular pathway with a self-sustained forward propagation of activity. 
Importantly, in addition to the propagation of the excitatory activity, we observe that the \textit{inhibitory} neurons' activity progresses from one group to the next. 
This emphasizes the pivotal role played by inhibitory units for the observed dynamics, which is clearly beyond simply balancing the network.

For our simulations, to describe the statistical \linebreak strength of the aforementioned connectivity motifs, we have defined an effective forward connectivity parameter $Q$, which is simply the ratio of the connection probabilities of the `targeted' vs. `non-targeted' neuron groups (or the ratio of corresponding synaptic weights, respectively), that modulates the amount of feedforward structure (see Materials and Methods).
For simplicity, we kept all these ratios equal. 
Note, however, that feedforward activity can be observed by changing only the weights or the connectivity probabilities separately (for a related observation, see \citet{Schaub2015}).
Using this simple setting, by varying $Q$ as our only parameter, we can alter the overall feedforward structure.
Note, $Q = 1$ corresponds to the case where the network is perfectly uniform. 
By increasing $Q$ to values larger than $1$ the level of feedforward structure increases.

The network simulated in Fig.~\ref{fig:2}a consists of 5 layers of neurons with a feedforward ratio of $Q = 2.6$.
To illustrate that increasing $Q$ indeed results in increased feedforward activity, we calculated the cross-covariance (Fig. \ref{fig:2}b) and the Pearson correlation coefficient (Fig. \ref{fig:2}c) between the firing patterns of the neurons, averaged over the different layers.

Fig.~\ref{fig:2}b shows the average cross-covariance functions within the same layer for the two conditions of $Q = 1$ and $Q = 2.6$, averaged over 10 realizations of the network.
To get a smooth estimate, we convolved the spike-train of every neuron with a Gaussian signal of standard deviation  $5ms$.
For every neuron pair, the convolved signal $f_i(t)$ was then used to calculate the pairwise cross-covariance $\phi_{ij}= \text{cov}[f_i(t+\tau),f_j(t)]$:
\begin{align}
\phi_{ij}(\tau)\approx \int[f_i(t+\tau) - \mu(f_i)][f_j(t) - \mu(f_j)]dt,
\end{align}
which we averaged over all neurons inside the same layer.
Here $\mu(\cdot)$ denotes the mean of the signal.
While there is no apparent temporal structure in the networks with $Q=1$, there is a clearly visible increased synchrony in the networks with high feedforward connectivity ratio $Q=2.6$, as indicated by the large peak at zero lag. 
Moreover, a second set of peaks appears at a lag of $\pm \delta$ corresponding to the repetition period of the firing, resulting from the imposed circular topology.

\begin{figure*}[tb!]
  \centering
  \includegraphics{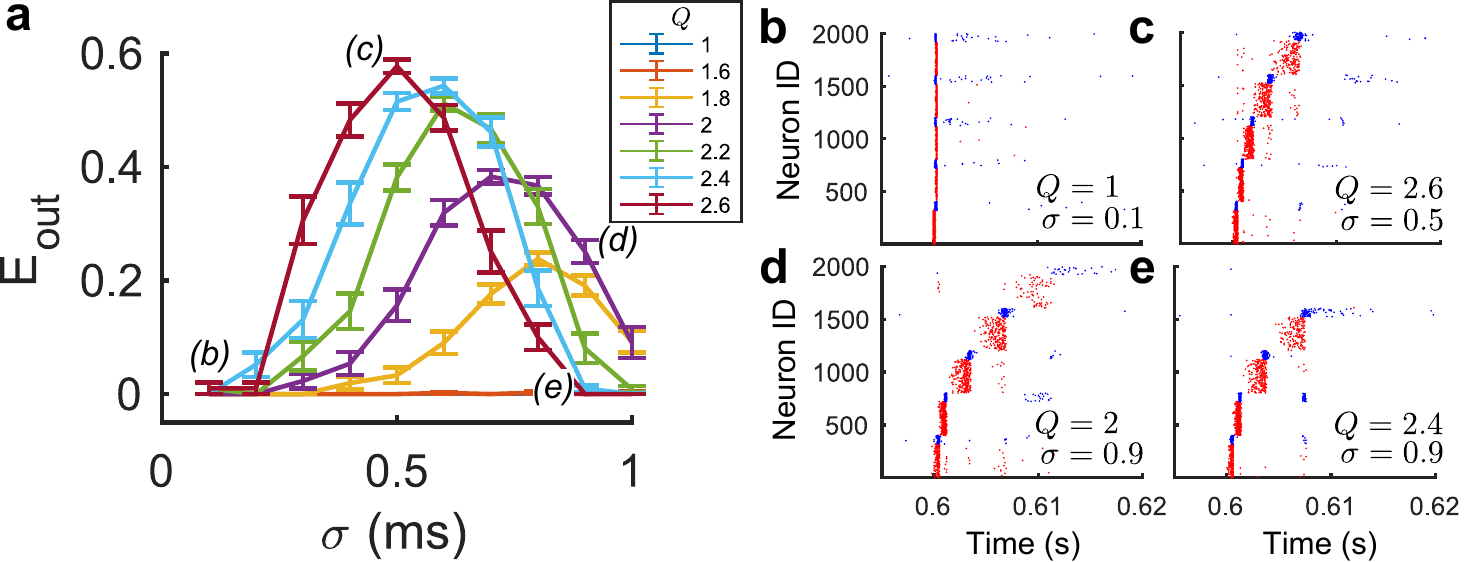}
  \caption{\textbf{Input-output characteristics of cross-coupled feedforward networks.} 
  \textbf{(a)} Plot of the efficacy of feedforward propagation, $E_{out}$ (see methods), as a function of how spread out the input signal is ($\sigma$) for different values of $Q$. Every simulation was repeated for 30 trials and error bars represent the standard error of the mean.
  \textbf{(b-e)} Examples for different levels of input spread $\sigma$ (ms) and $Q$ illustrating the different kinds of propagation possible with these networks: \textbf{(b)} all layers are activated (globally) by a pulse input, if the network is unstructured; \textbf{(c)} for sufficiently large $Q$, feedforward activity will be elicited; \textbf{(d-e)} however, depending on the parameter settings the feedforward activity may die out before reaching the last layer.
  }
  \label{fig:3}
\end{figure*}

To further investigate the tendency for each layer to fire in unison, we computed the Pearson correlation coefficient of the convolved spike-trains of all neuron pairs for varying levels of $Q$.
We plotted the average correlation coefficient within each group and layer in Fig.~\ref{fig:2}c. 
For each value of $Q$, 10 network realizations were simulated. 
As can be seen, the larger the value of $Q$, the more synchrony there is within groups.
This synchronous firing of groups is not decoupled but propagates along layers as can be seen in Fig.~\ref{fig:2}d. 
There we plotted the average cross-covariance of a layer relative to all other layers (c.f Fig.~\ref{fig:2}b).
As the regular shifts in the cross-covariance indicate, there is indeed a clear consecutive progression of activity from one layer to the next.
This shift is dependent on $Q$ in both amplitude (see below) and duration as seen in Fig.~\ref{fig:2}e which shows a plot of the time period for the signal to propagate around the network as a function of the feedforward connectivity parameter.
It is observed that increasing the value of $Q$, increases the value of the delay.
We can thus conclude that the ratio $Q$ directly influences the propagation of activity in ccFFNs, demonstrating that directed information transmission is possible without an imposed excitatory-to-excitatory pathway in the network.

\subsection{Assessing the input-output characteristics of cross-coupled feedforward networks.}
In the previous section we have seen that forward propagating activity can emerge in a ccFFN, provided the feedforward ratio $Q$ is sufficiently large enough.
In the following we examine this behavior in more detail to characterize the established feedforward circuits from an input-output viewpoint.

We consider a ccFFN circuit similar to above, but this time in a chain instead of a ring-like configuration, i.e., the last layer does not link back to the first layer, creating a clear input-output structure.
To maintain balance in the network, the input weights to the first excitatory group were shuffled, thereby removing the circular link back to group 1 (see Methods).
All other parameters were kept identical, but in order to delineate the effect of the input from spontaneous propagation, the net background drive was slightly reduced (see Methods).
Using this setup we can thus assess how a localized pulse at the first layer propagates towards subsequent layers, parametrically dependent on the feedforward connectivity ratio $Q$ (Fig.\ref{fig:3}).

We applied Gaussian impulses with varying temporal spread $\sigma$ to the first layer of excitatory neurons in the network and measured how the thus induced activity spread further to subsequent layers (see Fig.~\ref{fig:3}a).
The current injections occurred at $t = 600ms$ and the amplitude of current was normalized by sigma to ensure the net input was constant between simulations.

Nevertheless, as seen in Fig.~\ref{fig:3}b for an example impulse for a network with $Q = 1$ (no structure, fully random network), when the first excitatory group is stimulated to fire, the rest of the network fires in unison after a delay due to the large stimulation into one fifth of the excitatory neurons in the network.
To compensate for this effect, we employed a simple metric that is only nonzero if activity propagates from one group to the next, thereby ensuring that we do not pick up erroneous activity patterns.

Specifically the (feedforward) output energy $E_{out}$ at the last layer was calculated as follows:
\begin{enumerate}
    \item The group-averaged cross-covariance was calculated as done in Fig.~\ref{fig:2}b.
    \item To ensure feedforward activity propagation, we checked that $t_{peak5} > t_{peak4} > t_{peak3} > t_{peak2} > t_{peak1}$, otherwise we set $E_{out} = 0$.
        Note that this feature is primarily the reason why we see first a rise and then a fall in Fig.~\ref{fig:3}a for various values of $Q$, when varying $\sigma$:
        For very small values of sigma, the network can become globally active, such that activity peaks between different groups overlap (and are thus not ordered).
    \item If the above condition is satisfied, we calculated the net activity of the fifth group at $t_{peak5} \pm 5ms$ and divide that by the summed activity at $t_{peak1} \pm 5ms$ for the first group to attain $E_{out}$.
\end{enumerate}
As can be seen in Figure \ref{fig:3}a, by increasing $Q$ we can indeed trigger the ccFFN to propagate activity forward.
One could consider comparing this propagation activity with more traditional, excitatory driven feedforward networks.
However, as observed in numerical simulations (data not shown), this would require a major reduction in the excitatory-to-excitatory coupling, both in terms of the probability and weight of connections.
This is in agreement with previous reports \citep{Vogels2005} where embedding excitatory feedforward networks in balanced recurrent networks required the networks to be very sparse. 
Let us remark in this context again that our main point here has not been to construct a form of circuitry that provides a `superior' form of input-output mapping, but rather to show that there are multiple ways to implement such a mapping -- in our case, even without imposing a particular excitatory feedforward pathway.
\subsection{Disinhibitory feedforward networks}
The aforementioned cross-coupling of excitatory and inhibitory neurons is not the only arrangement possible to create feedforward activity driven by inhibitory units.
We noted numerous reports of long-range GABAergic neurons innervating predominantly or exclusively inhibitory neurons in downstream targets (\citet{Toth1993,Freund1988,Acsady2000,Gulyas1990,Rocamora1996,Freund1991,Freund1992,Melzer2017} but see also \citet{Pinto2006}). 
More recent studies have even shown how differential activation or suppression of specific subtypes of long-range inhibitory neurons (Parvalmbumin positive vs Somatostatin positive in different regions) can result in different motor system behaviors in the mouse \citep{Melzer2017}. 
From these reports, we sought to test if our ccFFN finding is general and should not be reduced to a single type of circuitry.
We present here a second network architecture which takes inspiration from the specificity of inhibitory neurons' connectivity patterns~\citep{Roux2015,Harris2015}.
In disinhibition motifs, certain subtypes of interneurons inhibit other interneurons which normally suppress connected excitatory neurons.
Such a disinhibition cascade can thus lead to an increase in firing rates in excitatory neurons which are normally suppressed. 
Although long range disinhibition reports are clear candidates for the proposed topology, we stress the architecture presented here does not impose spatial constraints. 

\begin{figure}[bt!]
  \centering
  \includegraphics{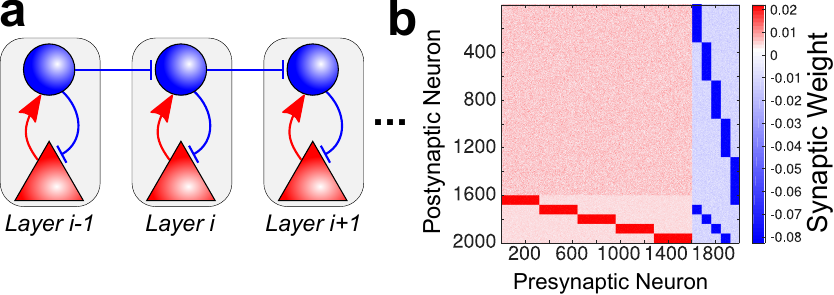}
  \caption{\textbf{Disinhibitory feedforward network.} 
  \textbf{(a)} Schematic of dFFN architecture, in which feedforward activity propagates by the cascading of several disinhibitory structural motifs. 
  For clarity, not all connections are shown in the schematic; importantly, excitatory to excitatory connections are randomly connected (see text).
  \textbf{(b)} Synaptic weight matrix of an example network with five groups. Note that there are connections between all types of neurons, as described in the Method section (\textbf{a} only emphasizes the pathways altered in the dFFN architecture). 
}
  \label{fig:4}
\end{figure}

Inspired by these experimental findings, we constructed a network model in which disinhibitory motifs guide the spiking activity.
We hereafter call this architecture a disinhibitory feedforward network (dFFN). 
A diagram of the wiring scheme for a dFFN is shown in Fig.~\ref{fig:4}a that highlights the key features of the network and does not include all connections.
For that, Fig.~\ref{fig:4}b shows a weight matrix instantiation with 5 groups (with the last group connecting to the first) which illustrates the full connectivity of dFFNs.

The functionality of this circuit can be explained conceptually as follows: 
(i) each layer comprises a functional group of excitatory and inhibitory neurons more strongly connected to each other than to the rest of the network; 
(ii) inhibitory neurons in one layer target preferentially the inhibitory neurons in the subsequent layer (disinhibition);
(iii) thus, when the activity in the preceding layer increases, the inhibitory neurons' activity in the next layer will decrease. 
(iv) This in turn allows the excitatory neurons in the next layer to increase their firing;
(v) after a short delay, the inhibitory neurons increase in activity again, as they receive input from the excitatory neurons in their layer, which have elevated activity as a result of their disinhibition. 
This eventually reduces the total firing back to baseline in the group -- however, not without the activity moving to the next group via disinhibition again; 
(vi) by cascading this motif, every upstream layer is activated and the information propagates.

\begin{figure*}[tb!]
  \centering
  \includegraphics{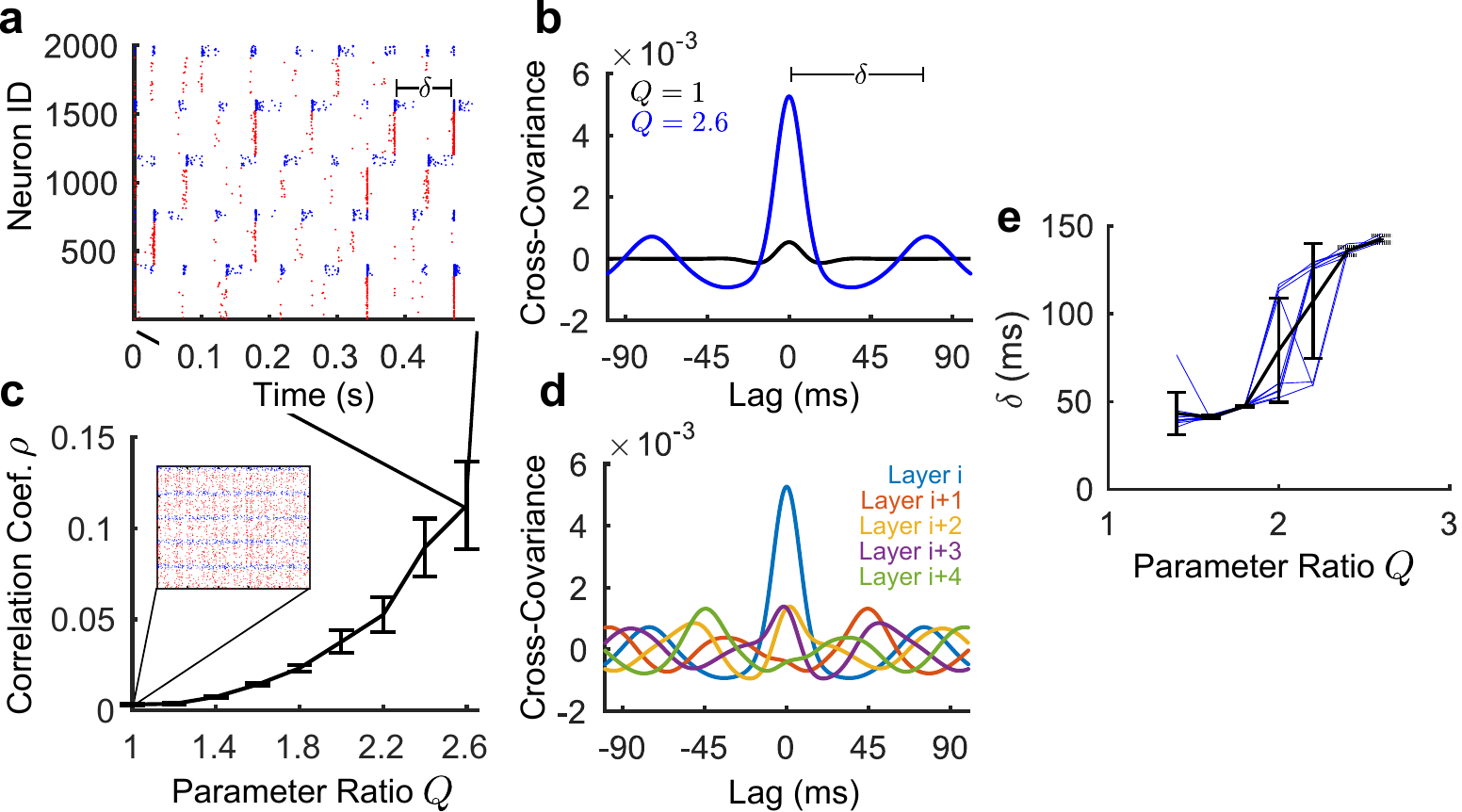}
  \caption{\textbf{Firing pattern characterization of disinhibitory feedforward network.} 
  \textbf{(a)} Example raster plot of a network with 5 layers showing forward propagation ($Q=2.6$). 
  The inhibitory neurons again display forward propagating activity. 
  Note that the final group connects back to the first (circular arrangement) and hence the activity propagates indefinitely.
  \textbf{(b)} Cross-covariance functions for the cases of $Q = 1$ and $Q = 2.6$ (see Fig.~\ref{fig:1}).
  \textbf{(c)} The average Pearson correlation coefficient within layers as a function of $Q$.
  The larger the feedforward ratio, $Q$, the greater the correlation of firing within layers.
  Error bars are standard deviation.      
  \textbf{(d)} 
  Average cross-covariance function between different layers (see also Fig. \ref{fig:1}). 
  Note that due to concurrently propagating multiple cascades, the cross-covariance between different groups display multiple peaks (see text).
 \textbf{(e)} Plot of the time period $\delta$ it takes the signal to propagate around the network as a function of $Q$.
      Note that for very small $Q$, the cross-covariance would have essentially no secondary peaks (see (b)).
      Thus we placed a threshold that the secondary peak needed to be at least half as large as the primary peak and only plotted $\delta$ values where that was satisfied.}
  \label{fig:5}
\end{figure*}

We implemented dFFNs with varying number of layers confirming that the above circuitry results in directed information propagation.
Once again we observe feedforward signal propagation for both the excitatory and inhibitory neurons (Fig.~\ref{fig:5}a). 
As before, to avoid boundary effects, we used a circular network layout in which the last group connects back to the first and hence the activity keeps propagating.

Similar to the ccFFNs, the level of feedforward structure was controlled by a parameter $Q$ that determines the ratio of connection probabilities and ratio of weights in the network to realize a dFFN (see Materials and Methods).
The displayed network in Fig.~\ref{fig:5}a corresponds to a 5-layer network with $Q = 2.6$.
When comparing the average cross-covariance function of such a network with uniformly connected networks $Q = 1$, we can again see an increased synchrony and a propagation of activity resulting from the dFFN architecture (Fig.~\ref{fig:5}b).
This is further demonstrated by the increasing Pearson correlation coefficient of neurons within layers as a function of $Q$ (Fig.~\ref{fig:5}c).
Finally, we plot again the average cross-covariance between different layers in Fig.~\ref{fig:5}d as well as the period duration of $\delta$ in Fig.~\ref{fig:5}e.
Although the same behavior is qualitatively seen of the signal propagating along the cascaded layers of the dFFN topology, this appears to be far less pronounced than in the ccFFN architecture.
The reason for this effect can be explained by inspecting the raster plot further (Fig.~\ref{fig:5}a).
Note that, in the ordered rastergram, the lowest group can fire again even though the cascade that emanated from it previously has not reached the top group yet.
For instance, the firing of a new cascade at layer $i$ and a previous cascade at layer $i+3$ can temporally overlap and this co-alignment results in the multiple peaks in the cross-covariance.
Hence, the network propagates multiple signals concurrently, or stated differently, the network's architecture is able to process multiple signals simultaneously effecting the cross-covariance measure.

\subsection{Assessing the input-output characteristics of disinhibitory feedforward networks.}
We performed a similar set of experiments as outlined for ccFFNs on disinhibitory networks where the circular topology was removed (see Methods).
Unlike for the ccFFN, for the disinhibitory architecture the results were less robust.

\begin{figure}[tb!]
  \centering
  \includegraphics[width=\columnwidth]{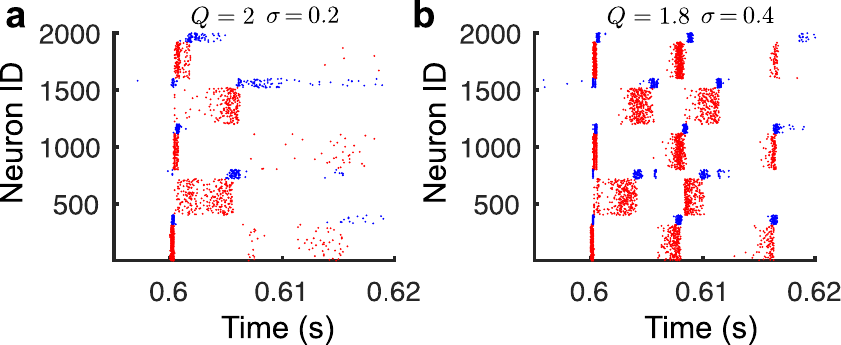}
  \caption{\textbf{Linear chain of disinhibitory feedforward circuits: overlapping cascades and interference.} 
      \textbf{(a)} Raster plot for a network with feedforward connectivity ratio $Q = 2$ and input with variance $\sigma = 0.2$ms (see text).
      Note that layers 3 and 5 are activated slightly after layer 1. 
      For this parameter setting the cascading behavior dies out quickly.
        \textbf{(b)} Raster plot for $Q = 1.8$ and $\sigma = 0.4$ms. 
  In contrast to the previous setting, the cascading behavior is sustained in this example. This is however not the case, in general.}
  \label{fig:6}
\end{figure}

Our simulations show that a strong impulse at $t =600ms$ can trigger a rapid successive firing of group 3 and group 5.
Likewise, the inhibitory neurons of groups 2 and 4 would be active in similar succession (see raster plots in Fig.~\ref{fig:6}a,b).
This phenomenon leads to the effect that propagation cascades can overlap, as is already evident from Fig.~\ref{fig:5}d.
A key driver for this behavior is that inhibitory neurons of even (odd) numbered groups get partially synchronized via spontaneous activity during the period where no input is applied, as can be seen when examining the membrane potentials in more detail (not shown).
The corresponding excitatory neurons inherit this partial synchronization via their coupling.
Now, when an impulse is applied, multiple cascades of activity may be triggered.
However, these cascades can interfere with each other through the global excitatory coupling in the network.
Such interference can lead to a breakdown of the cascade.

It is interesting to note that a signal may travel from input to output later in quite an uncommon sense: it is the lack of inhibition that `skips' subsequent layers and travels along the inhibitory connections, which in turn make the associated excitatory neurons prone to fire.
We think that by further adjusting the network, e.g., by introducing axonal delays and not forcing $Q$ to be equal throughout the network, more traditional feedforward patterns may be realized.

\section{Discussion}
We have demonstrated, via LIF network simulations, that one can construct networks in which feedforward activity is propagated even if connections between excitatory neurons are kept completely random. 
This is achieved by endowing inhibitory neurons with an active role in the feedforward propagation.

In this context the work of \citet{Ponzi2010} is worth mentioning, who observed heterogeneous (stochastic) cell assembly activity~\citep{Litwin-Kumar2012,Schaub2015} in networks with randomly connected inhibitory neurons.
Interestingly, as studied by \cite{Kopell2000}, inhibition patterns may also underpin long-range synchronization dynamics in gamma and beta frequencies. 
However, in these former studies the interplay of conduction delays and the network structure is a key element to be considered for the synchronization~\citep{Kopell2000}.
It would be interesting to investigate in how far such conduction delay effects may also be of relevance in the context of feedforward networks or related computational mechanisms.
More broadly, how inhibitory interactions may be utilized to implement reliably repeatable trajectories within neural networks, e.g., by shaping particular meta-stable cognitive states~\citep{Rabinovich2008}, appears be a promising avenue with plenty of room for future explorations.

Two circuit layouts were proposed demonstrating that both inhibitory and excitatory neurons can display feedforward dynamics simultaneously. 
We remark that the `layers' discussed within both of these network types should not be taken too literally. 
While within our wiring schemes certain subpopulations are targeted (statistically), there is still a (biased) all-to-all connectivity probability throughout.
Circuits like the ones discussed here, in which inhibitory neurons play a major role in directing information, might thus be found within one cortical column, for instance.

While long range projections in mammalian brains tend to be excitatory, topologies as discussed here might also be implemented by long range excitatory connections targeting local inhibitory neurons (see Fig.~\ref{fig:1}a). 
Moreover, long-range inhibitory projections have been increasingly reported in recent years \citep{Roux2015,Buzsaki2004,Alonso1982,Lee2014,Alonso1984,Freund1988, Tomioka2005,Jinno2007,Caputi2013,Gulyas2003,Toth1993,Freund1988,Acsady2000,Gulyas1990,Rocamora1996,Freund1991,Freund1992,Melzer2017}. 
However, we emphasize that our models are not dependent on projection lengths and thus are not limited to a particular bio-physical scenario by construction.
Moreover, the topologies we propose here also depend to some extent on a lack of connections, which make them more difficult to observe and report on experimentally.
The evidence is growing that neural networks could implement the herein described circuits, and our models prove that it is indeed possible to propagate information via specific inhibitory connections.
Whether such circuits' primary purpose would be to transmit information or control and gate signals is an exciting question to be considered in more detail in future investigations.

Our first circuit was inspired by the hippocampal architecture recently uncovered between CA3 and CA2 where the propagation of a signal between CA3 to CA2 pyramidal cells is governed by the amount of inhibition CA2 interneurons impose on their CA2 pyramidal cells~\citep{Nasrallah2015}.
Interestingly, there exist some further experimental indications that a similar mechanism to direct activity might be implemented in canonical cortical microcircuits~\citep{Pluta2015}. 
The traditional view of these ubiquitous circuits is that inputs from layer 4 (L4) drive layers 2/3 (L2/3) that then excites layer 5 (L5).
However, as Pluta and coworkers~\citep{Pluta2015} have shown, the picture is likely to be more intricate: 
in particular, it appears that L4 first suppresses L5 while driving L2/3, and only afterwards L5 shows elevated activity~\citep{Pluta2015} --- a finding that shows parallels to our proposed ccFFN mechanism. 
Moreover, a recent circuit reconstruction study of the rat entorhinal cortex identified a similar circuit at an axon-length scale via electron microscopy \citep{Schmidt2017}. 
\citet{Schmidt2017} observed that excitatory axons exhibit distance dependent targeting.
Specifically, an excitatory axon targeting both an inhibitory ($I_\text{target}$) and an excitatory target ($E_\text{target}$) will connect to $I_\text{target}$ earlier (shorter distances from the soma/axon-hilloc) than $E_\text{target}$ (larger distances).
Further, it is common that neuron $I_\text{target}$ connects to $E_\text{target}$, leading to a situation similar to that displayed in Fig.~\ref{fig:1}, albeit with the connection motif now realized on the axonal level.
The authors find that such spatially ordered synaptic contacts along the axon (first made with the interneurons, and only subsequently with the excitatory targets) lead to a temporal ordering of the activation of inhibitory and excitatory targets.
Numerical simulations showed that this enables the control of synchronized activity propagation and spike-timing \citep{Schmidt2017}.

The second circuit was inspired by the disinhibitory role of interneuron subtypes \citep{Pfeffer2013,Roux2015,Harris2015,Karnani2016} in addition to subtype specific plasticity rules \citep{Chen2015}.
Such discoveries and the advancement of connectomics in uncovering the diversity of neuronal cell types and their corresponding connectivity rules influenced our proposed wiring scheme of dFFNs.
For example, one of the key disinhibitory subtypes are the vasoactive intestinal peptide (VIP) neurons that have been shown to specifically target other inhibitory neurons but not themselves \citep{Pfeffer2013,Fu2014,Fu2015}.
More recently, their role in a behavioral task of a Go/No-Go task showed that when activating VIP neurons, the performance of the animal was enhanced  in contrary to activating the remaining inhibitory subtypes \citep{Kamigaki2017}.
This is complementary to other work that showed that differential activation or suppression of inhibitory neuronal subtypes results in differing motor system behaviors \citep{Melzer2017}. 
Indeed such control may be top-down to the circuit for different behaviors \citep{Zhang2014,Lee2017}, such as locomotion \citep{Lee2017a}, but our models here rely on the idea of concatenating such motifs to have a signal travel to multiple downstream targets consecutively, or even interlaced with more traditional feedforward architectures.
However, our numerical simulations showed that in a chain like linear configuration this architecture is less robust at displaying feedforward activity, when compared to ccFFNs.
Investigating these issues in more detail will be an interesting subject of future work.
We are encouraged to see that since our initial submission~\citet{Murray2017} presented an all-inhibitory circuit model of the stratium.
\citet{Murray2017} demonstrated, via rate models and spiking models, how recurrently connected inhibitory neurons arranged in a circular topology with \textit{weakened} connections can result in a propagation of patterns from one unit to the next. 
Moreover, it was shown that by changing the excitatory input drive to the network, the speed of propagation can be adjusted by an order of magnitude \citep{Murray2017}.
 
Overall, the recently discovered diversity of interneurons suggests that they play a much more vital role in neuronal network dynamics than simply acting as a balancing device for the network. 
We believe that such new experimental discoveries call for a reassessment of the role of inhibitory neurons in models or neuronal circuits, and encourage scholars to assign them more active and functional roles. 
Here we have demonstrated how such a functional role could be shaped in feedforward networks, but possible roles clearly go beyond this.

\section{Materials and Methods}
Simulations were performed in MATLAB (2012b or later) and code can be found at \url{github.com/CellAssembly/inhibitory-feedforward}.

\subsection{Model of spiking neurons}
To illustrate our ideas, we have used leaky-integrate-and-fire (LIF) networks, stylized models of neural networks, which act like pulse-coupled oscillators.
Using a time step of $0.1$ms we numerically integrated the non-dimensionalized membrane potential of each neuron, which evolved according to:
\begin{equation}\label{eq:LIF_model}
  \dfrac{d V_i(t)}{dt} = \dfrac{1}{\tau_m}(\mu_i - V_i(t)) + \sum_{j}W_{ij}g^{E/I}_j(t), 
\end{equation}
with a firing threshold of $1$ and a reset potential of $0$.
All networks comprised $N=2000$ units, with an excitatory to inhibitory neuron ratio of $4:1$ ($1600$ excitatory, $400$ inhibitory).
The input terms $\mu_i$ were chosen uniformly in the interval $[1.1, 1.2]$ for excitatory neurons, and in the interval $[1, 1.05]$ for inhibitory neurons. 
Membrane time constants for excitatory and inhibitory neurons were set to $\tau_m = 15$~ms and $\tau_m = 10$~ms, respectively, and the refractory period was fixed at $5$~ms for both excitatory and inhibitory neurons. 
Note that although the constant input term is supra-threshold, balanced inputs guaranteed an average sub-threshold membrane potential~\citep{Litwin-Kumar2012,Schaub2015}.

In the model, the network coupling is captured by the sum in~\eqref{eq:LIF_model}, which describes the input to neuron $i$ from all other neurons in the network.
Here $W_{ij}$ denotes the weight of the connection from neuron $j$ to neuron $i$ ($W_{ij}=0$ if there is no connection).
After a presynaptic spike of neuron $j$, the synaptic inputs $g^{E/I}_j(t)$ are increased step-wise ($g_j^{E/I}\rightarrow g_j^{E/I} +1$) instantaneously, and then decay exponentially according to:

\begin{equation}\label{eq:synapse}
    \tau_{E/I} \dfrac{d g_j^{E/I}}{dt} = - g_j^{E/I}(t),
\end{equation}
with time constants $\tau_E =3$ ms for an excitatory interaction, and $\tau_I =2$ ms if the presynaptic neuron is inhibitory.
For all networks described in the following, the total connection-strength per neuron was kept equivalent to an unstructured, balanced network displaying asynchronous activity, with $p_{EI}=p_{IE}=p_{II}=0.5$, $p_{EE}=0.2$, $w_{EI} = w_{II} = -0.042$,  $w_{IE}= 0.0105$, and $w_{EE} = 0.022$.
Here, $p$ and $w$ stand for the connection probability and connection weight, respectively. 
The first subscript denotes the destination and the second superscript denotes the origin of the synaptic connection, and $E$,$I$ stand for an excitatory or inhibitory neuron, respectively.

\subsection{Cross-coupled feedforward networks (ccFFN)}
To construct ccFFNs we kept excitatory-to-excitatory and inhibitory-to-inhibitory connections uniform as outlined above.
We divided the network into layers consisting of both inhibitory and excitatory units and connected as outlined in Fig.~\ref{fig:1}b,c:
Excitatory neurons in layer $i$ are statistically biased to target the inhibitory neurons in their own layer with a weight ratio $W_{IE} = w^{i}_{IE} / w^{not[i]}_{IE}$, compared to the inhibitory neurons in the rest of the network.
Similarly, the inhibitory neurons within a layer $i$ target the excitatory neurons in the \textit{next} layer $i+1$, more weakly according to the ratio $W_{EI} = (w^{i+1}_{EI} / w^{not[i+1]}_{EI})^{-1}$.
In addition to modifying the weights, we control the analogous ratio of connections probabilities $R_{IE}$ and $R_{EI}$. 
Note that $W_{EI}, R_{EI}$ are defined with an inverse ratio, i.e., a higher ratio means a \textit{weaker} targeting corresponding to a stronger feedforward structure.
To modulate the embedded feedforward level in the networks, we can thus vary the ratios $W_{IE}, W_{EI}, R_{IE}, R_{EI}$, while keeping the average weights and number of connections constant.

For assessing the input output characterizations, we removed the circular topology while keeping a balanced network.
To this end, the input links to the first layer where shuffled randomly, thereby breaking the circular topology.
This ensured the networks propagated activity due to our proposed architecture and  not due to unbalanced excitatory-inhibitory ratios.
For the plots in Fig.~\ref{fig:3}, the input drive to excitatory neurons was slightly reduced to $[1.0, 1.05]$, which reduced spontaneous activation and thereby allowed us to observe the effects caused by the input pulses more clearly. 
Note that all other parameters were kept constant.

\subsection{Networks driven by disinhibitory structure (dFFN)}
In dFFN networks, excitatory-to-excitatory connections remain again uniform.
Every layer in this network is composed of groups of excitatory and inhibitory units as shown in Fig.~\ref{fig:2}a,b.
Similar to above, we can control the imposed feedforward level with the ratio parameters $W_{IE} = w^{i}_{IE} / w^{not[i]}_{IE}$, $W_{EI} = w^{i}_{EI} / w^{not[i]}_{EI}$, and $W_{II} = w^{i+1}_{II} / w^{not[i+1]}_{II}$ or the analogous ratios of connections probabilities $R_{IE}, R_{EI}, R_{II}$.
Again, for simplicity we set all six parameters equal to $Q$ and vary them concurrently.

Similar to the ccFFN topology, the input-output relationship required the breaking up of the circular topology while maintaining network balance.
Here, the links contacting the first inhibitory group were shuffled to maintain a balanced input to the first layer of inhibitory neurons. 
To delineate the effects of the input more clearly, the input drive to excitatory nodes was again reduced to $[1.0, 1.05]$ for the results shown in Fig.~\ref{fig:6}.
Otherwise, the network parameters were not changed.

\begin{acknowledgements}
We are thankful for discussions with Christof Koch, Costas Anastassiou, and Mauricio Barahona and comments from Jean-Charles Delvenne and Renaud Lambiotte.
YNB wishes to thank the Allen Institute founders, P. G. Allen and J. Allen, for their vision, encouragement and support.
Most of this work has been performed while MTS was at the Universit\'e catholique de Louvain.
MTS acknowledges support from the ARC and the Belgium network DYSCO (Dynamical Systems, Control and Optimisation) and an F.S.R. fellowship of the Universit\'e catholique de Louvain.
MTS has received funding from the European Union’s Horizon 2020 research and innovation programme under the Marie Sklodowska-Curie grant agreement No 702410.
\end{acknowledgements}

\bibliographystyle{spbasic}      
\bibliography{references}

\end{document}